# pyBioSig: optimizing group discrimination using genetic algorithms for biosignature discovery

Frederico G. C. Arnoldi[1,a], Rodrigo F. Rodrigues[1], Celio L. Silva[1]
[1]Departmento de Bioquímica e Imunologia, Faculdade de Medicina de Ribeirão Preto, Universidade de São Paulo, Ribeirão Preto, SP, CEP 14049-900, Brasil.
[a] email: fgcarnoldi@gmail.com

**ABSTRACT**

In medical sciences, a biomarker is "a characteristic that is objectively measured and evaluated as an indicator of normal biological processes, pathogenic processes, or pharmacologic responses to a therapeutic intervention". Molecular experiments are providing rapid and systematic approaches to search for biomarkers, but because single-molecule biomarkers have shown a disappointing lack of robustness for clinical diagnosis, researchers have begun searching for distinctive sets of molecules, called "biosignatures". However, the most popular statistics are not appropriate for their identification, and the number of possible biosignatures to be tested is frequently intractable. In the present work, we developed a "multivariate filter" using genetic algorithms (GA) as a feature (gene) selector to optimize a measure of intra-group cohesion and inter-group dispersion. This method was implemented using Python and R (pyBioSig, available at https://github.com/fredgca/pybiosig under LGPL) and can be manipulated via graphical interface or Python scripts. Using it, we were able to identify putative biosignatures composed by just a few genes and capable of recovering multiple groups simultaneously in a hierarchical clustering, even ones that were not recovered using the whole transcriptome, within a feasible length of time using a personal computer. Our results allowed us to conclude that using GA to optimize our new intra-group cohesion and inter-group dispersion measure is a clear, effective, and computationally feasible strategy for the identification of putative "omical" biosignatures that could support discrimination among multiple groups simultaneously.

**INTRODUCTION**

Complex systems are present in any human issue, but their intrinsic structure prevents them from being intuitive, and specific methods are required for interpreting them. For the scientific community, one of the valid strategies is to reduce their complexity. This method, frequently known as feature or dimensionality reduction, tries to find single features or small sets of features that reflect the whole scenario, or summarize all features in a smaller number of compounded new ones (Saeys et al, 2007; Hilario and Kalousis, 2008; Li et al, 2008; Ringnér, 2008;).

In biological systems, these distinctive features are generally named "biomarkers" and can be found and made useful in any area of the natural sciences, but medical research is seeking them most intensely (Gustaw-Rothenberg et al, 2010; Parida and Kaufmann, 2010; Pedraza et al, 2010). In the medical context, a biomarker is "a characteristic that is objectively measured and evaluated as an indicator of normal biological processes, pathogenic processes, or pharmacologic responses to a therapeutic intervention" (Group., 2001). Although many characteristics have the potential to be used as biomarkers, molecular experiments are providing rapid and systematic approaches to search for them (Liu et al, 2005; Agranoff et al, 2006; Kussmann et al, 2006; Gonzalez-Juarrero et al, 2009; Nagaraj, 2009).

Medical biomarkers could help researchers to understand biochemical (Fernandes et al, 2004) or immune responses (Querec et al, 2009), classify cancer in types or levels (Brawer, 2000; Liu et al, 2005; Pedraza et al, 2010), identify the progression and stages of any other disease (Saeys et al, 2007; Gonzalez-Juarrero et al, 2009; Gustaw-Rothenberg et al, 2010), evaluate effectiveness of and patient responsiveness to a treatment (Parida and Kaufmann, 2010), and predict behaviors (Fernandes et al, 2004), among other applications.

As single-molecule biomarkers for clinical diagnosis have shown a disappointing lack of robustness (Brawer, 2000), researchers have been looking for distinctive sets of molecules, called "biosignatures" (Liu et al, 2005; Agranoff et al, 2006; Gonzalez-Juarrero et al, 2009; Pedraza et al, 2010). Simple calculations of fold changes, t and F tests, do not measure classification accuracy and do not explore gene combinations (Xiong et al, 2001). Otherwise, distance between samples can be used with multivariate data and is appropriate for classifications. A brief discussion of it can be found in the paper by D'haeseleer (D'haeseleer, 2005).

Distances can help with handling many variables at once but do not help us select them. On the other hand, trying all possible gene combinations is frequently impossible (Hilario and Kalousis, 2008). Genetic algorithms (GA) is a heuristic search technique that performs optimizations by mimicking the process of natural evolution (Goldberg, 1989) and can be used as a feature (gene) selector. The paper by Ooi and Tan (2003) provided a short description of the method and how it can be modeled for gene selection.

In the present work, we developed and implemented a "multivariate filter" (Saeys et al, 2007) that optimizes the discrimination of any number of groups, using genetic algorithms as the feature (gene) selector and a simple measure of intra-group cohesion and inter-group dispersion as objective function. We believe that the method presented and its implementation would be useful for researchers' efforts to identify putative biosignatures.

**METHODS**
**Software framework**

We developed the pyBioSig program using Python (Python Programming Language) as the main programming language and the R package (The R Project for Statistical Computing) as the statistical framework. The pyEvolve library (Christian, 2009) was used as Genetic Algorithms framework, pyGTK (PyGTK: GTK+ for Python) for interface development, and the RPy library (RPy: A simple and efficient access to R from Python) for Python/R communication. We employed the pvclust library for R (Suzuki and Shimodaira, 2006) to perform hierarchical clustering and to calculate bootstrap and approximately unbiased p-values (Suzuki and Shimodaira, 2006) of each generated cluster.

**Genetic algorithm parameters**

We used genetic algorithm as a gene (feature) selector. The parameters used in all analyses were adapted from (Peng et al, 2003), with RouletteWheel as the selector, a mutation ratio of 1%, a crossover ratio of 80%, a population with 40 individuals, and 50000 generations.

The objective function to be optimized by our GA was the minimization of n largest and mean distances among members from the same group and the simultaneous maximization of the n smallest and mean distances among members from different groups, as summarized by the formula

$$1000 - \left(\frac{wmax + F.wmean}{bmin + F.bmean}\right) - \text{size penalty}$$

where wmax and wmean represent the sum of the n largest and the mean distances among members from the same group, respectively, and bmin and bmean represent the sum of the n smallest and the mean distances among members from different groups, respectively. "F", a weight for the mean distances on the equation, and "n" were set by default to 0.2 and 1, respectively. Euclidean distance was used as the default distance measure in all analyses.

To obtain biosignatures with a reduced number of probes, we also included a linear penalty function that decreases with the number of genes included. This penalty is canceled when the biosignature reaches a desired range and is set to the maximum penalty when the biosignature size becomes smaller than the minimum size.

**Benchmarking**

The ability of our method to selectively recover informative probes was tested over artificial datasets, composed of different numbers of known informative, uninformative and noisy probes. Informative probes were generated following a normal (s.d.: 0.3) or uniform (boundaries: ±0.6 from the midpoint) distribution with different means or midpoints, respectively, for each or one group. Uninformative probes were generated following a normal distribution (s.d.: 0.75) with the same mean for all samples. Values from noisy probes were generated following a normal distribution (s.d.: 0.25) with different means for at least one group with members assigned randomly. Means or midpoints ranged from 3 to 15 in the integer set and were randomly assigned. The gain in power of classification was evaluated by multidimensional scaling and hierarchical clustering, comparing the results of our specific filter against the results of Anova (a nonspecific filter). Bootstrap and approximately unbiased p-values (Suzuki and Shimodaira, 2006) were used to evaluate the robustness of the clusters produced and of the probe set selected.

The same evaluations were also performed with the following published microarray data: GSE4511 (Maurer et al, 2005), GSE5500 (Bisping et al, 2006), GSE5429 (Fernandes et al, 2004). With the GSE5429 dataset, in order to evaluate the tendency of our method to overfit, we also performed a round o cross-validation in the form of leave-one-out, testing the classification of all odd samples using a K-NN classifier, using two neighbors for classification (Larrañaga et al, 2006).

**Preprocessing and nonspecific filtering of published microarray data**

Bioconductor (Gentleman et al, 2004) was used to perform all preprocessing and nonspecific filtering tasks. Microarray data were loaded using the Affy package (Gautier et al, 2004). Background adjustment, quantile normalization and median polish summarization were performed in all microarray data using GCRMA (Wu et al, 2004). Those probes with log2 signal lower than 2.5 in more than 75% of the samples or interquartile range lower than 0.5 were eliminated using the Genefilter package. Probes with p-value greater than 0.01 (without any correction for multiple testing) in an Anova analysis were also removed.

**RESULTS**

Our method was implemented using Python and R in a software that we called pyBioSig, available at https://github.com/fredgca/pybiosig under LGPL. The parameters of a single analysis using pyBioSig can be set via our graphical interface, but batch analyses can be also performed using Python scripts in order to optimize the value of these different parameters, like biosignature size and objective function weights.

To evaluate and test the pyBioSig method, we created some data sets with 40,000 artificial probes and four groups of 15 samples each. From these 40,000 probes, 4 (0.01%) had coherent information for all four groups, 4 (0.01%) had coherent information for just one group and no information for the others, 4,000 (10%) had noisy information that clusters samples in groups different from those we have originally set, and 35,992 (89.9%) probes had no information, i.e., all samples had values coming from the same distribution. In one of these data sets, for example, the nonspecific filter (Anova) selected 388 probes, including all 8 informative, 344 uninformative and 36 noisy probes.

The pyBioSig parameters were set as described in Material and Methods to search for the most distinctive probe set with a size ranging from 6 to 12 probes. In 30 independent searches over three artificial data sets, like that mentioned above, created independently, our method selected informative probes predominantly (figure 1). In numbers, 96.6% of the putative biosignatures contained none or just a single uninformative/noisy probe. These searches, using a single personal computer with an i3 Intel processor, took a median of 6.7 hours with an interquartile range of 0.22 hours (13 minutes), but reasonable results could be obtained in less than 90 minutes.

As we can see in figure 2, a representative solution from our artificial data, the probe set from pyBioSig showed visible improvement over the nonspecifically filtered set. By hierarchical clustering, the main evidence of this improvement was the correct clustering of all samples, which did not happen with the nonspecifically filtered data. Higher bootstrap and approximate approximately unbiased p-values [31] for the known groups indicated more stable clusters. We were also able to observe that distances among members from the same group were smaller and more uniform, whereas distances among members from different groups were larger in the pyBioSig result than in the nonspecifically filtered. With a multidimensional scaling plot, we could observe that the groups in our probe set were more compact, but different groups were more dispersed from each other, as expected.

Once we had obtained evidence that our methodology and its implementation were effective in retrieving probes with information for discrimination among groups selectively, we tested pyBioSig with real microarray data. With the data sets GSE4511 (Maurer et al, 2005), GSE5500 (Bisping et al, 2006), GSE5429 (Fernandes et al, 2004) and unpublished data from our laboratory (Rodrigues RF, et al. unpublished results), we were able to obtain results as successful as with our artificial data set. Figure 3 shows an example of the improvements with the data set GSE5429, which has eight groups with four samples each, representing mRNA extracted from the hippocampus of eight mouse strains with different aggressive behavior. With hierarchical clustering, all groups were correctly clustered with higher bootstrap and approximately unbiased p-values. In the multidimensional scaling plot, we also observed more compact groups but more distant from each other. The results of our cross-validation, described in Material and Methodos, classified all testing samples correctly, showing no tendency to overfit the biosignatures to the training data.

**DISCUSSION**

The biomarker quest seeks simple signals, mainly at the molecular level, to understand or answer questions like "Which type of cancer does patient X have?", "Will BCG protect this child/population against tuberculosis?", "In which stage is the disorder?". The ideal would be finding one cheap, easy, and objective measure that would robustly answer the question or predict the outcome. However, taken together that single-feature markers have disappointing lack of robustness (Brawer, 2000) and most questions are multi-class problems, i.e., they are not yes/no questions, we need a set of markers or a signature to have confident results.

Considering that we can include or not include a molecule in a signature, having x candidate molecules, we have $2^x$ possible biosignatures. Frequently, a nonspecific filtering of "omical" data leaves us with something around 200 candidate molecules, which means $2^{200}$ possible signatures. Considering that our universe has a history of less than $2^{60}$ seconds, it is clear that we cannot try all possible combinations and that we need an intelligent approach.

In the present work, we developed a method based on genetic algorithms that seeks probe sets of predetermined size that improve group discrimination. The method was implemented as a computer program that we called pyBioSig. As shown in our results, starting with an artificial data set with 388 probes, the result of a nonspecific filtering employing Anova, we were able to find probe sets predominantly containing only informative probes. This result indicates that our method is effective in identifying and selecting informative probes within a feasible length of time with an inexpensive personal computer.

Applying the same strategy to real microarray data, we had similar results, making groups internally more cohesive but externally more dispersed. We highlight the results obtained with the GSE5429 data set. This data set has eight groups, representing mRNA extracted from the hippocampus of eight mouse strains with different aggressive behavior. By applying a nonspecific filter, 646 probes were selected, but hierarchical clustering using Euclidean distance and average linkage were not able to cluster samples according to their strains. Likewise, multidimensional scaling revealed very confused groups. However, when we applied the pyBiosig method, we were left with 11 probes that robustly clustered all samples according to their strains. Of 50 randomly created probe sets of the same size from the nonspecifically filtered data, none of them clustered all samples correctly. Altogether, these results indicate that our method and its implementation are effective in making groups more distinguishable.

The method with the current objective function, described in the Methods section, tries to minimize the sum of the n largest and the mean distances among members from the same group while maximizing the sum of the n smallest and the mean distances among members from different groups. The rationale behind this objective function was to be a mid-point between k-means, that minimizes within-group distances, and SVM, that tries to maximize between-group distances. Another important consideration was that the method at all, could be applied to small datasets.

In this approach, the largest distances among members from the same group and the smallest ones among members from different groups are the most important values to be optimized. However, when groups have overlapping members, taking these measures into account demands the optimization of the parameter "n". As the mean distance embraces all individuals, it would optimize general group cohesion and dispersion and make the method simpler in the latter case. The "F" and "n" factors allow researchers to adapt the objective function to different data sets by giving different weights to each component of the objective function and/or by changing the number of extreme values to be optimized.

As a multivariate filter, it just selects features according to the predefined function. In order to have a diagnostic tool, it is necessary to couple it with a discriminant method, like hierarchical clustering, nearest-neighbour, decision trees or SVMs (Larrañaga et al, 2006). Multidimensional scaling, although imprecise, gives a very intuitive output.

Although our method showed interesting results, to have reasonable biosignature candidates, reliable data from good experimental design are necessary and, as for any prediction, they need confirmation before can be considered true biosignatures

(Koulman et al, 2009). It is noteworthy that complete discrimination among different groups is not always possible, as it also depends heavily on the data structure.

Every test was performed with microarray data for biochemical discrimination. However, in our opinion, there is no reason to believe that the same method will not work with different "omical" data and for different purposes. In fact, it would work with any data set from any research field when the purpose is to find those variables that best discriminate among different groups.

Besides presenting a new method for putative biosignature identification, as far we know, pyBioSig is the first free and open-source tool for the purpose that allows simple and comfortable analysis via a graphical interface but also allows the use of scripts for batch analyses. On the one hand, Python scripts offer flexibility and allow researchers to test the effect of setting different parameters more efficiently. Also, as GA is a stochastic method, it is convenient to try different runs with the same initial parameters (Russell and Norvig, 2009). On the other hand, as our objective function does not require understanding of complex linear algebra and the evolutionary process is intuitive to most researchers of natural sciences, we believe that unlike previous works (Ooi and Tan, 2003; Peng et al, 2003; Liu et al, 2005), with the simple interface provided, we are also providing a powerful and effective tool accessible even to researchers with lower computer expertise.

**CONCLUSIONS**

Our results allow us to conclude that using GA to minimize the largest and mean distances among members from the same group while maximizing the smallest and mean distances among members from different groups is an effective and efficient strategy for the identification of putative molecular biosignatures that allow discrimination among multiple groups. It is also clear and computationally feasible and, in theory, can be used to seek putative biosignatures for any number of groups simultaneously, even from small datasets.

**ACKNOWLEDGEMENTS**
This work was developed with the financial support of CNPq. We would like to thanks Luana Silva Soares for reviewing our manuscript.

**REFERENCES**
Agranoff D, Fernandez-Reyes D, Papadopoulos MC, Rojas SA, et al. (2006) Identification of diagnostic markers for tuberculosis by proteomic fingerprinting of serum. Lancet 368: 1012-1021

Bisping E, Ikeda S, Kong SW, Tarnavski O, et al. (2006) Gata4 is required for maintenance of postnatal cardiac function and protection from pressure overload-induced heart failure. Proc Natl Acad Sci U S A 103: 14471-14476

Brawer MK (2000) Prostate-specific antigen. Semin Surg Oncol 18: 3-9

Christian SP (2009) Pyevolve: a Python open-source framework for genetic algorithms. SIGEVOlution 4: 12-20

D'haeseleer P (2005) How does gene expression clustering work? Nat Biotechnol 23: 1499-1501

Fernandes C, Paya-Cano JL, Sluyter F, D'Souza U, et al. (2004) Hippocampal gene expression profiling across eight mouse inbred strains: towards understanding the molecular basis for behaviour. Eur J Neurosci 19: 2576-2582

Gautier L, Cope L, Bolstad BM, Irizarry RA (2004) affy--analysis of Affymetrix GeneChip data at the probe level. Bioinformatics 20: 307-315

Gentleman RC, Carey VJ, Bates DM, Bolstad B, et al. (2004) Bioconductor: open software development for computational biology and bioinformatics. Genome Biol 5: R80

Goldberg DE (1989) Genetic Algorithms in Search, Optimization and Machine Learning, 3rd edn. Addison-Wesley Longman Publishing Co, Boston.

Gonzalez-Juarrero M, Kingry LC, Ordway DJ, Henao-Tamayo M, et al. (2009) Immune response to Mycobacterium tuberculosis and identification of molecular markers of disease. Am J Respir Cell Mol Biol 40: 398-409

Group. BDW (2001) Biomarkers and surrogate endpoints: preferred definitions and conceptual framework. Clin Pharmacol Ther 69: 89-95

Gustaw-Rothenberg K, Lerner A, Bonda DJ, Lee HG, et al. (2010) Biomarkers in Alzheimer's disease: past, present and future. Biomark Med 4: 15-26

Hilario M, Kalousis A (2008) Approaches to dimensionality reduction in proteomic biomarker studies. Brief Bioinform 9: 102-118

Koulman A, Lane GA, Harrison SJ, Volmer DA (2009) From differentiating metabolites to biomarkers. Anal Bioanal Chem 394: 663-670

Kussmann M, Raymond F, Affolter M (2006) OMICS-driven biomarker discovery in nutrition and health. J Biotechnol 124: 758-787

Larrañaga P, Calvo B, Santana R, Bielza C, et al. (2006) Machine learning in bioinformatics. Brief Bioinform 7: 86-112

Li GZ, Bu HL, Yang MQ, Zeng XQ, et al. (2008) Selecting subsets of newly extracted features from PCA and PLS in microarray data analysis. BMC Genomics 9 Suppl 2: S24

Liu JJ, Cutler G, Li W, Pan Z, et al. (2005) Multiclass cancer classification and biomarker discovery using GA-based algorithms. Bioinformatics 21: 2691-2697

Maurer LM, Yohannes E, Bondurant SS, Radmacher M, et al. (2005) pH regulates genes for flagellar motility, catabolism, and oxidative stress in Escherichia coli K-12. J Bacteriol 187: 304-319

Nagaraj NS (2009) Evolving 'omics' technologies for diagnostics of head and neck cancer. Brief Funct Genomic Proteomic 8: 49-59

Ooi CH and Tan P (2003) Genetic algorithms applied to multi-class prediction for the analysis of gene expression data. Bioinformatics 19: 37-44

Parida SK, Kaufmann SH (2010) The quest for biomarkers in tuberculosis. Drug Discov Today 15: 148-157

Pedraza V, Gomez-Capilla JA, Escaramis G, Gomez C, et al. (2010) Gene expression signatures in breast cancer distinguish phenotype characteristics, histologic subtypes, and tumor invasiveness. Cancer 116: 486-496

Peng S, Xu Q, Ling XB, Peng X, et al. (2003) Molecular classification of cancer types from microarray data using the combination of genetic algorithms and support vector machines. FEBS Lett 555: 358-362

PyGTK: GTK+ for Python - http://www.pygtk.org.

Python Programming Language - http://www.python.org.

Querec TD, Akondy RS, Lee EK, Cao W, et al. (2009) Systems biology approach predicts immunogenicity of the yellow fever vaccine in humans. Nat Immunol 10: 116-125

Ringnér M (2008) What is principal component analysis? Nat Biotechnol 26: 303-304

RPy: A simple and efficient access to R from Python - http://rpy.sourceforge.net.

Russell S and Norvig P (2009) Artificial Intelligence: A Modern Approach, 3rd edn. Prentice Hall, Englewood Cliffs.

Saeys Y, Inza I, Larrañaga P (2007) A review of feature selection techniques in bioinformatics. Bioinformatics 23: 2507-2517

Suzuki R and Shimodaira H (2006) Pvclust: an R package for assessing the uncertainty in hierarchical clustering. Bioinformatics 22: 1540-1542

The R Project for Statistical Computing - http://www.r-project.org.

Wu Z, Irizarry RA, Gentleman R, Martinez-Murillo F, et al. (2004) A Model-Based Background Adjustment for Oligonucleotide Expression Arrays. Journal of the American Statistical Association 99: 909-917

Xiong M, Fang X, Zhao J (2001) Biomarker identification by feature wrappers. Genome Res 11: 1878-1887

**Figures**

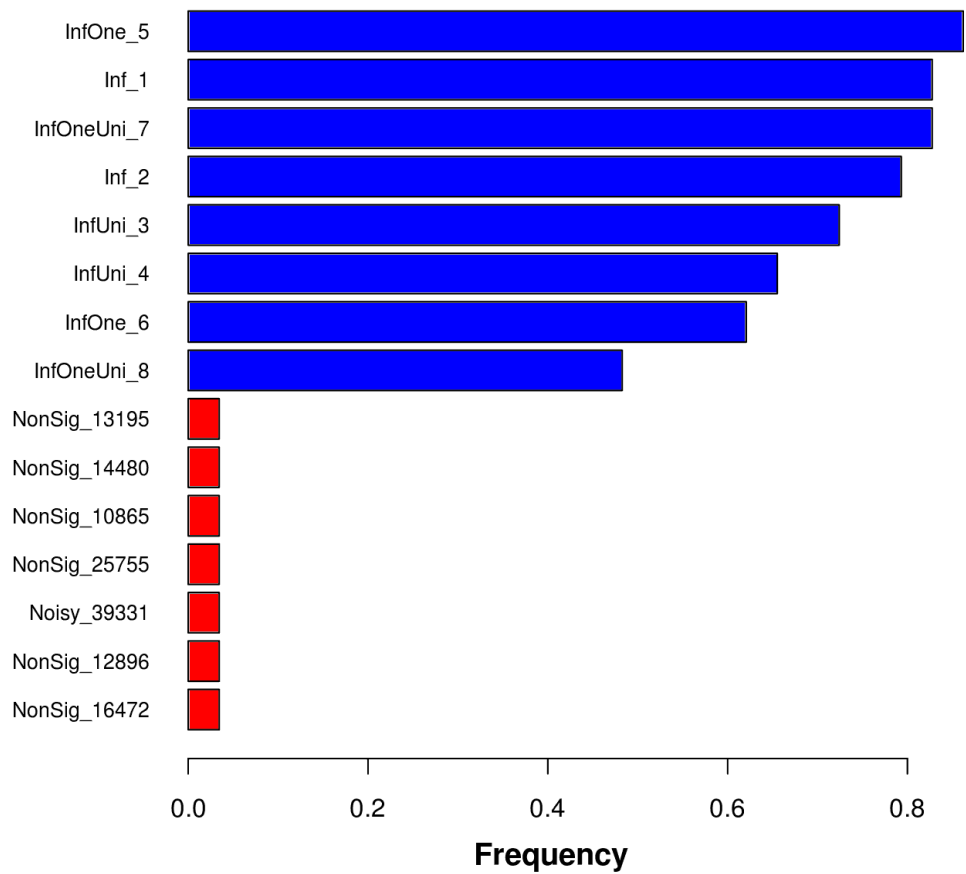

**Figure 1.** Frequency of the different probes selected in 30 independent searches for putative biosignatures over an artificial data set. Blue bars represent the frequency of selected informative probes, whereas red ones represent the frequency of selected uninformative or noisy probes.

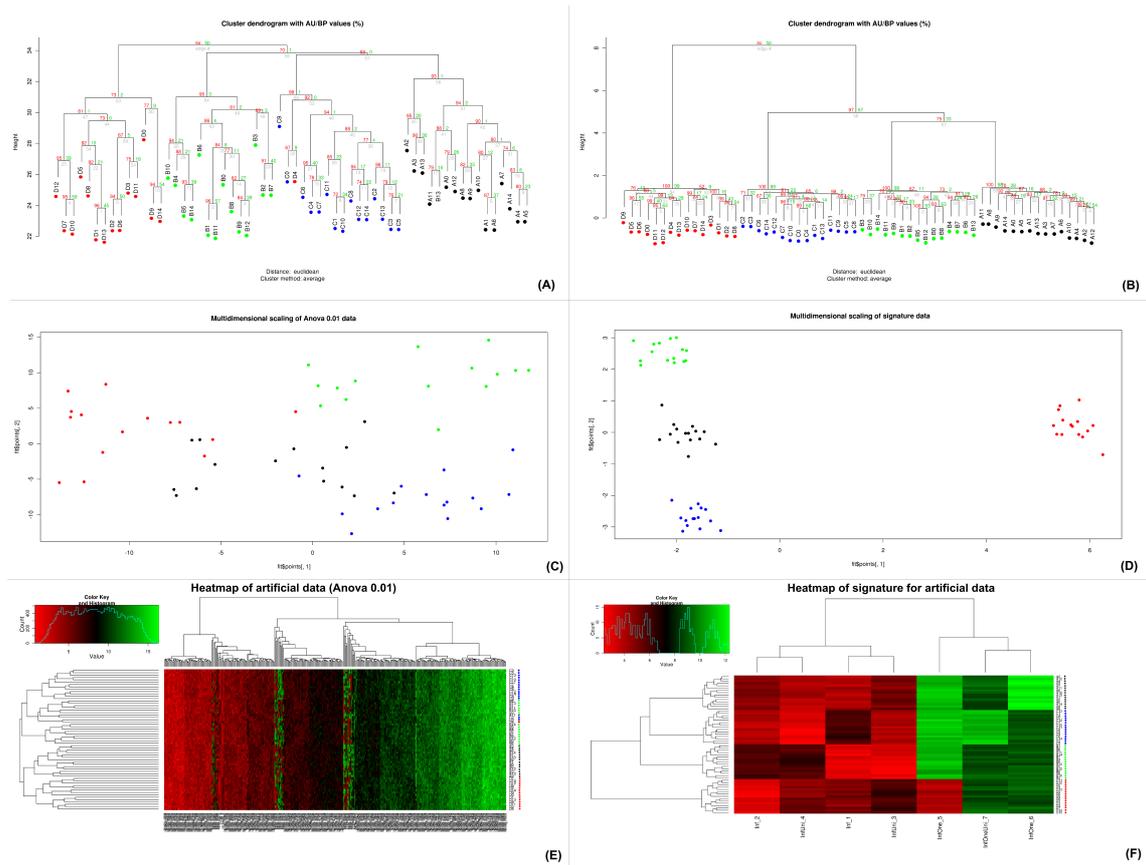

**Figure 2.** Comparison of the probes in the biosignature generated using pyBioSig software and those from a nonspecific filter of the artificial data set. The figures on the right are analyses of biosignature data, while the ones on the left are from the nonspecific filter. **A** and **B** are the hierarchical clusterings using Euclidean distance and clustering by the mean distance (UPGMA); the green values above each cluster represent bootstrap support, while the red ones are approximately unbiased p-values. **C** and **D** are multidimensional scalings using Euclidean distance, with the color of points representing samples from the same group as in **A** and **B**. **E** and **F** are heatmaps of the resulting probes, with rows and columns ordered by hierarchical clustering.

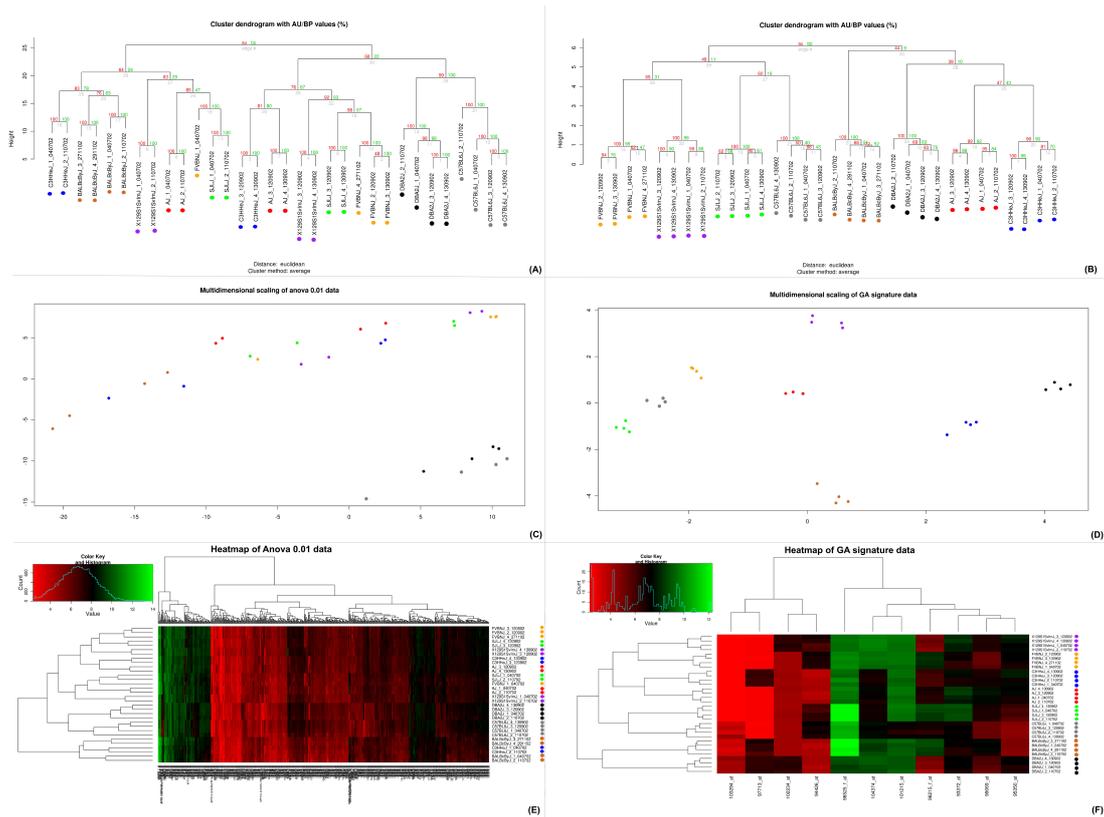

**Figure 3.** Comparison of the probes in the biosignature generated using pyBioSig software and those from the nonspecific filter of the GSE5429 data set. The right-hand figures are analyses of biosignature data, while the left-hand ones are from the nonspecific filter. **A** and **B** are the hierarchical clusterings using Euclidean distance and clustering by the mean distance (UPGMA); the green values above each cluster represent bootstrap support, while the red ones are approximately unbiased p-values. **C** and **D** are multidimensional scalings using Euclidean distance, with the color of points representing samples from the same group as in **A** and **B**. **E** and **F** are heatmaps of the resulting probes, with rows and columns ordered by hierarchical clustering.